\documentclass[11pt]{article}

\usepackage{moriond,epsfig}

\def\Journal#1#2#3#4{{#1} {\bf #2}, #3 (#4)}

\def\NIMA{{\em Nucl. Instrum. Methods} A}

\def\PLB{{\em Phys. Lett.}  B}

\def\PRL{\em Phys. Rev. Lett.}

\def\PRD{{\em Phys. Rev.} D}

\def\be{\begin{equation}}

\def\ee{\end{equation}}

\def\bea{\begin{eqnarray}}

\def\eea{\end{eqnarray}}

\usepackage{relsize}

\def\babar{\mbox{\slshape B\kern-0.1em{\smaller A}\kern-0.1em B\kern-0.1em{\smaller A\kern-0.2em R}}}

\begin{document}

\vspace*{4cm}

\title{B DECAYS TO HADRONIC STATES WITH CHARM/CHARMONIUM IN BABAR}

\author{ G. Vuagnin\\ (for the \babar\ Collaboration) }

\address{INFN Sezione di Trieste, Area di Ricerca, Padriciano 99,\\
34012 Trieste, ITALY}

\maketitle\abstracts{
In this paper the studies of three different B decays to hadronic states are 
presented. These results are based on 1999-2003 dataset collected by the \babar\ experiment at the PEP-II $e^+ e^-$ storage ring at the Stanford Linear Accelerator Center.
The measurements are the hadronic branching fraction of 
$B^+ \rightarrow  J/\Psi p \bar{\Lambda}$,  $B^0 \rightarrow  J/\Psi p \bar{p}$,  $B^0 \rightarrow  D^{*\pm} D^{\mp}$, 
 and the direct $CP$-asymmetry in 
$B^0 \rightarrow  D^0(CP) K^-$ channels.}

\section{Introduction}
  The decay of $B$ mesons to open charm and charmonium  provides an excellent
  laboratory for the study of hadronic $B$ decays. With about $88$ millions 
  of $B$ pairs, 
  the \babar\ experiment has collected a sample of data that enable to test models of B decay
   in more modes and with greater precision than ever before. In this note we present some  
   examples. In section~\ref{charmonium}  a test on non-relativistic QCD
  is described. In the next 
  section a measurement of branching fraction and measurements of time-integrated asymmetries in channels suitable for 
 CP violation  study are  presented.
The \babar\ detector is described in detail elsewhere~\cite{babar}.
Charge conjugation is implied throughout this note.

\section{B Decays To Charmonium States}
\label{charmonium}
 The inclusive production of charmonium mesons in $B$ decay at the 
$\Upsilon(4S)$ shows an excess of $J/\Psi$ mesons at low center-of-mass 
momentum $ p_{CM}$~\cite{balest,aubert,schrenk}, when compared to distributions 
predicted by non-relativistic QCD calculations~\cite{beneke}.
Some  hypothesis have been 
proposed to explain the sources of the excess:  an
intrinsic~\cite{chang} charm component 
of the B,  the production of an $s\bar{d}g$ hybrid~\cite{eilam} in conjunction with a $J/\Psi$, or the  possibility that the excess comes 
 from decays of the type $B \rightarrow J/\Psi$ baryon anti-baryon~\cite{brodsky}.
 The rate of these decays could be enhanced by the intermediate production of 
an exotic state allowed by QCD but not yet observed. In this section  we 
present the searches for the decays $B^+ \rightarrow J/\Psi p \bar{\Lambda}$ 
and $B^0 \rightarrow J/\Psi p \bar{p}$ as a test of this last hypothesis.

The reconstruction of $B^+ \rightarrow J/\Psi  p \bar{\Lambda}$  candidates is done  by combining
$J/\Psi$, proton, and $\Lambda$ candidates. $J/\Psi$ are reconstructed 
in the $e^+ e^{-} $  or  $\mu^+ \mu^-$ final states.
The selection of good proton candidates is one of the key element of the 
analysis.
The separation of low momentum protons from kaons is done by 
a likelihood method that uses energy deposited and Cherenkov angle measurements. 
At a typical momentum
of $300$MeV/c, the selection efficiency  is greater
than $98 \%$ with a kaon misidentification probability less than $1 \%$.
The $\Lambda$ is reconstructed from a proton and an oppositely charged track, 
assumed to be a pion. 
$B^0 \rightarrow  J/\Psi p \bar{p}$  candidates are formed from $J/\Psi $  candidates
and an oppositely-charged pair of proton candidates.
We used the kinematic variables $\Delta E$ and $m_{ES}$~\cite{babar} to 
characterize B 
candidates.
The analysis region considered is defined by 
$5.2 < m_{ES} < 5.3$GeV/c$^2$  and 
$-0.10 < \Delta E < 0.25$GeV ($B^+$ candidates) and
$-0.25 < \Delta E < 0.25$GeV ($B^0$ candidates).
For signal events, $\langle \Delta E \rangle \approx 0$ 
and $\langle m_{ES} \rangle \approx M_B$. 
We define a signal region as an ellipse with semi-axes proportional to the resolutions 
$\sigma_m$ and $\sigma_E$ estimated, from simulated data,
to be 3.1 MeV/c$^2$ and 6.5 MeV, respectively, for $B^+ \rightarrow J/\Psi p \bar{\Lambda}$, and 
2.7MeV/c$^2$ and 5.5MeV  for $B^0 \rightarrow J/\Psi p \bar{p}$.
We use simulated $B^+ \rightarrow  J/\Psi p \bar{\Lambda}$ and $B^0 \rightarrow  J/\Psi p \bar{p}$ events to estimate the selection efficiency that is $0.049 \pm 0.009$ for the charged channel and $0.184 \pm 0.024$ for the neutral one. We 
have studied the accuracy of the simulation of the detector response
by comparing data and simulated background events in
samples similar to the final selection.
 The expected background in the signal ellipse is extrapolated  from the number of
 candidates outside the ellipse in the analysis region considered.  
For  $B^+ \rightarrow  J/\Psi p \bar{\Lambda}$ we have 39 candidates in the analysis region implying an expected background of  $0.21 \pm 014$. We observe four candidates in the signal ellipse.
 The probability of  observing $\ge 4$ candidates when expecting $0.21\pm 0.14$ is $2.5 \times 10^{-4}$. To interpret this result as a $B^+$ branching fraction ${\cal B}$,
we undertake a Bayesian analysis with a uniform prior above zero. We
define the likelihood for ${\cal B}$ as the probability of observing exactly
four events, including uncertainties on the expected
background, signal efficiency, secondary branching fractions, and
number of $\Upsilon(4S)$ decays ($(88.9 \pm 1.0)\times 10^6$).
The result is
${\cal B}(B^+ \rightarrow  J/\Psi p \Lambda) = 11.6 ^{+8.5}_{-5.6} \times 10^{-6}$~\cite{chris}, 
where the uncertainty includes both statistical and systematic components. 
We similarly obtain a  
90\% CL upper limit of $26 \times 10^{-6}$.
For  the Cabibbo suppressed mode $B^0 \rightarrow  J/\Psi p \bar{p}$ we followed the same procedure. There are 126 events outside the signal ellipse.
The expected background is of $0.64 \pm 0.17$  and one event has been  found inside the ellipse. We obtain 
${\cal B}(B^0 \rightarrow  J/\Psi p \bar{p}) < 1.9 \times 10^{-6}$ (90\% CL). This limit is dominated by
statistical uncertainty. 
Neither final state makes a significant
contribution to the 
observed excess of $J/\Psi$ mesons in inclusive $B$ decay.

\section{B Decays To Open Charm States}
\label{charm}
In this section we describe the analysis of two channels interesting for CP 
violation studies. 

The measurements of the parameter $\sin (2\beta)$ using the quark process 
$b \rightarrow c \bar{c}s$ have shown that $CP$ is
violated in the neutral $B$-meson system~\cite{BabarSin2b,BelleSin2b}, consistently with the Standard Model (SM) expectation~\cite{CKMWorkshop}.
In order to search for additional sources of $CP$ violation from new physics processes, 
different quark decays such as a $b \rightarrow c\bar{c}d$ must be examined.
We describe here the measurements of branching fraction and of time-integrated 
$CP$-asymmetry in  $B \to D^{*\pm}D^{\mp}$. 
$B$ decays to
$D^{*+}D^{-}$ and $D^{*-}D^{+}$ are selected 
via full reconstruction of the decay products.  
The $D^{*+}$  is reconstructed in its decay to $D^0 \pi^+$, where the $D^0$ subsequently
decays to one of the four modes
$K^{-}\pi^{+}$,  $K^{-}\pi^{+}\pi^0$, $K^{-}\pi^{+}\pi^{-}\pi^{+}$, or $K^0_S\pi^{+}\pi^{-}$.
The $D^-$ is reconstructed
in its decays to $K^{+}\pi^{-}\pi^{-}$ and to $K^0_S\pi^{-}$.
B candidate selection is based on a likelihood which includes all  measured $D^{\pm}$, $D^0$
mass values and the  $D^*-D$ mass difference.
Candidates are then characterized by the kinematic variables $\Delta E$
 and $m_{ES}$. The signal region in $\Delta E$ is defined to be $|\Delta E| < 18$MeV. 
According to Monte Carlo simulations,
the width of this region corresponds to approximately 
twice the signal resolution.  
$B \to D^{*\pm}D^{\mp}$ candidates in the region $5.27 < m_{ES} < 5.30 $Gev/c$^2$ and 
$|\Delta E| < 18$ MeV  are used to extract signal events.
A sideband, defined as $5.20 < m_{ES} < 5.27 $GeV/c$^2$ and $|\Delta E| 
< 18$ MeV, and a ``large sideband'', defined as $5.20 < m_{ES} < 5.27 $Gev/c$^2$ and 
$|\Delta E| < 200$MeV, are used to extract various background parameters.  The total
numbers of selected events in the signal region, the sideband, and the large sideband 
are 197, 461, and 5187, respectively.
We use an unbinned extended maximum likelihood fit to the $m_{ES}$ distribution  to extract the number
 of signal events above background as well as the time-integrated $CP$ asymmetry, defined as:
\begin{equation}
\mathcal{A} = \frac{N_{D^{*+}D^{-}} - N_{D^{*-}D^{+}}}{N_{D^{*+}D^{-}} + N_{D^{*-}D^{+}}}.
\end{equation}
The $m_{ES}$ distribution for the simultaneous fit to
all the selected events is described by Gaussian distributions for the $D^{*+}D^
{-}$ and $D^{*-}D^{+}$
signals, an ARGUS threshold function~\cite{argus}, and a Gaussian distribution to describe a 
small "peaking'' background estimated to be of $12 \pm 8$ events, from studies performed with both data and Monte Carlo simulations.
There are a total of four free parameters in the nominal fit: the shape and normalization of the background ARGUS function (2), the total 
$B \to D^{*\pm}D^{\mp}$ signal yield (1), and the $CP$ asymmetry $\mathcal{A}$ (1).
We use a Monte Carlo simulation of the BaBar detector to
determine the reconstruction efficiencies, that range from 6\% to 18\%   
depending on the $D$ decay modes. 
From these efficiencies and the total number of recorded $B\bar{B}$ pairs, 
and assuming the $\Upsilon(4S)$ decaying equally in $B^+B^-$ and $B^0 \bar{B^0}$ we determine
the branching fraction to be:
$$\mathcal{B}(B \to D^{*\pm}D^{\mp}) = (8.8 \pm 1.0{\rm (stat.)} \pm 1.3{\rm (syst.)}) \times 10^{-4}.$$
\begin{figure}
\centering
\epsfig{figure=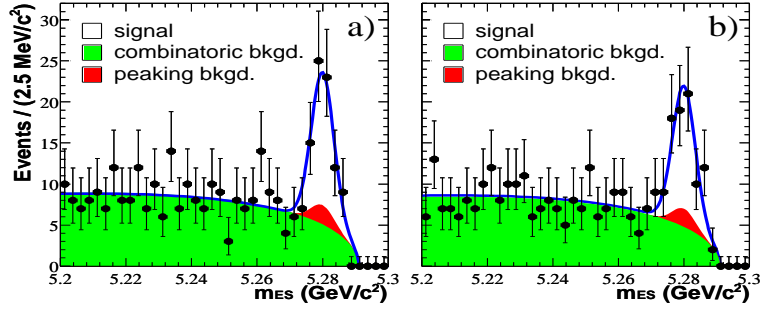,width=10cm,height=4cm}
\caption{The $m_{ES}$ distributions of a)
$B \to D^{*-}D^{+}$
and b) $B \to D^{*+}D^{-}$
candidates with $|\Delta E| < 18$MeV.
The fit includes Gaussian distributions to model the signal and 
a small peaking background component, and an ARGUS
function to model the combinatoric background shape.
\label{fig}}
\end{figure}
The total 
systematic uncertainty from all considered sources is $14.5\%$. 
The fitted value for $\mathcal{A}$ is 
$$\mathcal{A} = -0.03 \pm 0.11{\rm (stat.)} \pm 0.05{\rm (syst.)}.$$
Systematic uncertainties on $\mathcal{A}$ are dominated by potential differences
 in the reconstruction efficiencies 
of positively and negatively charged tracks (0.04), and by uncertainty in the $m_{ES}$
resolution for $B \to D^{*\pm}D^{\mp}$ signal events (0.03).

 A theoretically clean measurement of the angle $\gamma$ can be obtained  from
 the study of $B^- \rightarrow D^{(*)0}K^{*-}$ decay by reconstructing the $D^0$ meson 
 into Cabibbo allowed CP eigenstates and double Cabibbo suppressed decays~\cite{gronau,atwood}.
 We will describe here a first step through this analysis describing the measurement of the direct CP asymmetry defined as:
\begin{equation}
\mathcal{A_{CP}} = \frac{ {\cal B}(B^- \rightarrow D^{0}_{CP}K^{-}) - {\cal B}(B^+ \rightarrow D^{0}_{CP}K^{+})} { {\cal B}(B^- \rightarrow D^{0}_{CP}K^{-}) + {\cal B}(B^+ \rightarrow D^{0}_{CP}K^{+})}
\end{equation}
where $D^{0}_{CP}$ is a  $D^{0}$ meson reconstructed in either the Cabibbo allowed $K^+ K^-$ $CP$ final state or in the Cabibbo soppressed $\pi^+ \pi^-$ $CP$ mode.
 The analysis procedure for the two channels is similar but not identical.
 For the $K^+ K^-$ mode  the signal is extracted using a unbinned maximum likelihood fit 
 to the variables $\Delta E$, $m_{ES}$ and the kaon ID probability of the prompt track in the final state. For the $\pi^+ \pi^-$ mode the $m_{ES}$ variable is replaced by the $D^0$ invariant mass because of a possible dangerous contribution to this channel 
from the non resonant $ B^- \rightarrow K^{-} \pi^- \pi^+$ decay. 
The variable $m_{ES}$ has in fact the same property for the signal and this background while $D^0$ mass has respectively peaking or flat distribution in the two cases. The shapes of signal and background distribution are 
 determined by off-resonance data, Monte Carlo simulation and data control samples.
The measured direct $CP$-asymmetry is $0.17 \pm 0.23^{+0.09}_{-0.07}$  for $B^{\pm} \rightarrow D^{0}(K^+ K^-)K^{\pm}$ and $-0.44 \pm 0.34 \pm 0.06$ for $B^{\pm} \rightarrow D^{0}(K^+ K^-)K^{\pm}$.
The dominant sources of uncertainties are signal and background parametrization, particle ID and detector asymmetry.

\section{Conclusion}
 We have measured the branching fraction of  $B^+ \rightarrow  J/\Psi p \bar{\Lambda}$
 and $B^0 \rightarrow  J/\Psi p \bar{p} $ and we can conclude that these channels 
 could not be responsible of the observed excess of $J/\Psi$ mesons in inclusive 
 $B$ decay. We have shown the results of the measurement of the branching fraction 
 and time integrated $CP$-asymmetry  in the channel $B \to D^{*\pm}D^{\mp}$ 
 which is examined to analyse different modes measuring $\sin (2\beta)$, with the goal of  understand CP violation as well as penguin contributions thoroughly. 
 We have then shown the 
 measurement of direct $CP$ asymmetries in the channels  $B^{\pm} \rightarrow D^{0}(K^+ K ^-, \pi^+ \pi^-)K^{\pm}$ which will allow in the future, with a data sample few times 
  the current one, the measurement of the angle $\gamma$ of the Unitarity Triangle of 
 the CKM mixing matrix.


\section*{References}
\begin{thebibliography}{99}

\bibitem{babar}BABAR Collaboration, B. Aubert {\it et al.}, \Journal{\NIMA}{479}{117}{2002}.

\bibitem{balest}CLEO Collab., R. Balest {\it et al.}, \Journal{\PRD}{52}{2661}{1995}; S. Anderson {\it et al.}, \Journal{\PRL}{89}{282001}{2002}.

\bibitem{aubert}BABAR Collab., B. Aubert {it et al.}, \Journal{\PRD}{67}{032002}{2003}.

\bibitem{schrenk}S. Schrenk, in {\it Proceedings of the 30th International Conference 
on High Energy Physics}, edited by C.S. Lim and T. Yamanaka (Osaka, Japan, 2000), Vol. 2, p. 839. 
\bibitem{beneke}M. Beneke, G.A. Schuler, and S. Wolf, \Journal{\PRD}{62}{034004}{2000}.
 
\bibitem{chang}C.-H.V. Chang and W.-S. Hou, \Journal{\PRD}{64}{071501}{2001}.

\bibitem{eilam}G. Eilam, M. Ladisa, and Y.-D. Yang, \Journal{\PRD}{65}{037504}{2002}.

\bibitem{brodsky}S.J. Brodsky and F.S. Navarra, \Journal{\PLB}{411}{152}{1997}.


\bibitem{chris}BABAR Collab., B. Aubert {it et al.}, hep-ex/0303036 Submitted to {\PRL}

\bibitem{BabarSin2b}
BaBar Collaboration, B.~Aubert {\it et al.}, \Journal{\PRL}{89}{201802}{2002}.

\bibitem{BelleSin2b}
BELLE Collaboration, K.~Abe {\em et al.}, \Journal{\PRD}{66}{071102}{2002}.

\bibitem{CKMWorkshop} See, \textit{e.g.}, F.~Gilman, K.~Kleinknecht, and B.~Renk, \Journal{\PRD}{56}{010001}{2002}, and references therein.

\bibitem{argus}ARGUS Collab., H. Albrecht {it et al.}, Z. Phys. {\bf C48 } 543 (1990).
\bibitem{justin}BABAR Collab., B. Aubert {it et al.}, hep-ex/0303004 Submitted to {\PRL}

\bibitem{gronau} M. Gronau and D. Wyler, \Journal{\PLB}{265}{172}{1991}.
M. Gronau and D. London, \Journal{\PLB}{253}{483}{1991}.

\bibitem{atwood}D. Atwood, I.Dunietz and A. Soni, \Journal{\PRL}{78}{3257}{1997}.
\end{thebibliography}
\end{document}